\documentclass[reprint,prd,nofootinbib,preprintnumbers]{revtex4}
 \usepackage{graphicx,psfrag,epsfig,epsf,colordvi,slashed}
 \def\lsim{\mathrel{\vcenter{\hbox{$<$}\nointerlineskip\hbox{$\sim$}}}}

\begin{document}
\preprint{MAN/HEP/2012/14}
\title{Minimal Radiative Neutrino Mass Mechanism for Inverse Seesaw Models}
\author {\bf  P. S. Bhupal Dev and Apostolos Pilaftsis\\}
%
\affiliation{\vspace{2mm}
Consortium for Fundamental Physics, School of Physics and Astronomy,
University of Manchester, Manchester, M13 9PL, United Kingdom.} 

\begin{abstract}
\vspace{0.5cm}
\centerline{\bf ABSTRACT}
\vspace{1mm}
\noindent
We study  a minimal one-loop radiative mechanism  for generating small
Majorana neutrino masses in  inverse seesaw extensions of the Standard
Model with two singlet fermions per family. 
The new feature of this radiative mechanism is that the one-loop
induced   left-handed   neutrino  mass   matrix   is  {\em   directly}
proportional  to   the  Majorana  mass  matrix   of  the  right-handed
neutrinos. This is a very economical scenario without necessitating  the 
existence of non-standard scalar or gauge fields.
\end{abstract}

\maketitle

\section{Introduction}

The understanding of the  extra-ordinary smallness of neutrino masses,
along with their observed large mixings~\cite{vureview} remains one of
the major puzzles in the Standard  Model (SM), and so a potential portal to
New Physics.  The simplest theoretical scenario which may explain this
neutrino  puzzle would be  to introduce  singlet Majorana  masses that
break the global  $(B-L)$-symmetry of the SM. Within  the SM, this can
be parameterized  through the non-renormalizable  dimension-5 operator
due      to      Weinberg~\cite{weinberg}:      $\lambda_{ij}\,\left(L^{\sf T}_i\Phi\right)\left(
L_j^{\sf T}\Phi\right)/\Lambda$, where $L_i=(\nu_i,\ell_i)_L^{\sf T}$ (with $i=e,\mu,\tau$) is the
SU(2)$_L$  lepton  doublet,  $\Phi=(\phi^+,\phi^0)^{\sf T}$  is the  SM  Higgs
doublet, and $\Lambda$ is  an effective mass scale of the New Physics.  
After electroweak
symmetry breaking via the vacuum  expectation value (VEV) of the Higgs
field, $\langle \Phi\rangle  = (0,v)^{\sf T}$, the SU(2)$_L$-doublet neutrinos
$\nu_{iL}$   receive   a   non-zero    mass   matrix   of   the   form
$(M_{\nu_L})_{ij} = \lambda_{ij} v^2/\Lambda$.

There exist  three tree-level realizations~\cite{type1,type2,type3} of
this   effective  dimension-5   operator  using   only  renormalizable
interactions.   The   simplest  one,   widely  known  as   the  type-I
seesaw~\cite{type1},  requires the  extension of  the SM  by  a number
$n_R$  singlet   fermions,  which  are  usually  taken   to  be  heavy
right-handed   (RH)  neutrinos  $\nu_{\alpha   R}$  (with   $\alpha  =
1,2,\dots,n_R$).   After integrating  out the  heavy  singlet neutrinos
$\nu_{\alpha R}$, we  obtain Weinberg's effective dimension-5 operator
mentioned above. Specifically, the  neutrino masses and mixings may be
deduced from the Yukawa Lagrangian:
\begin{equation}
-{\cal L}_Y\ =\ (y_\nu)_{i\alpha}\, \bar L_i \Phi\,
\nu_{\alpha R}\: +\: \frac{1}{2}\bar\nu_{\alpha R}^C
(M_R)_{\alpha\beta} \nu_{\beta R}\ +\ {\rm  H.c.}~, 
\end{equation}
which in  turn implies  the following $(3+n_R)\times  (3+n_R)$ complex
symmetric matrix in the flavor basis $\{(\nu_{iL})^C,\nu_{\alpha R}\}$:
\begin{eqnarray}
{\cal M}_\nu\ =\ \left(\begin{array}{cc}
{\bf 0} & M_D \\
M_D^{\sf T} & M_R
\end{array}\right)\; ,
\label{eq:Mnutype1}
\end{eqnarray}
where  $M_D=y_\nu v$ is  the Dirac  mass matrix  for the  neutrinos and
$M_R$   is  the   $(B-L)$-breaking   Majorana  mass   matrix  of   the
RH neutrinos.       In     the     usual      seesaw     approximation
$(M_{D})_{i\alpha}/(M_{R})_{\alpha\beta}\ll   1$,   the   physical   light
neutrino masses are given by
\begin{eqnarray}
M_{\nu_L}\ \simeq\ -\,M_D M_R^{-1} M_D^{\sf T}\; .
\label{eq:type1}
\end{eqnarray}
Evidently, the seesaw scale $M_R$  is correlated with the magnitude of
the  Yukawa coupling $y_\nu$.   For the  experimentally observed  sub-eV light
neutrinos~\cite{vureview},  $M_R$ is  usually constrained  to  be much
larger  than  the electroweak  scale $m_W$,  unless  $y_\nu$  is very  small
($\lsim 10^{-6}$) [cf.~(\ref{eq:type1})] or there are cancellations in
the mass  matrix structure given  by (\ref{eq:type1}) due  to specific
flavor symmetric patterns of $y_\nu$ and $M_R$~\cite{canc}.

An interesting realization for the seesaw scale to be in the TeV range
is  the  so-called   inverse  seesaw  model~\cite{inverse},  where  in
addition  to RH  neutrinos  $\{\nu_{\alpha R}\}$,  another  set of  SM
singlet fermions  $\{S_{\rho L}\}$ are introduced. Here  we consider a
symmetric extension  of the SM  with three pairs of  singlet neutrinos
(i.e. $\alpha,\rho=1,2,3$). One of the salient features of the inverse
seesaw mechanism is that the  small neutrino masses are generated by a
{\it small}  lepton-number breaking  $3\times 3$ mass  matrix $\mu_S$.
In contrast to the type-I seesaw case, the mass matrix of the observed
left-handed  neutrinos  $\left(\nu_{e,\mu,\tau}\right)_L$ vanishes  in
the limit  of lepton-number conservation:  $\mu_S \to {\bf  0}$.  More
explicitly,  the neutrino Yukawa  sector of  a general  inverse seesaw
model is described by the Lagrangian
\begin{eqnarray}
-{\cal L}_Y &=& (y_\nu)_{i\alpha}\, \bar L_i\Phi\, \nu_{\alpha R}\:
+\: \bar{S}_{\rho L}\, (M_N)_{\rho\alpha} \nu_{\alpha R} 
+\ \frac{1}{2}\bar{\nu}^C_{\alpha R} (\mu_R)_{\alpha\beta}
\nu_{\beta R}\: +\: \frac{1}{2}\bar{S}_{\rho L}
(\mu_S)_{\rho\lambda} S^C_{\lambda L}\: +\ {\rm H.c.}~, 
\end{eqnarray}
which gives rise to the following $9\times 9$ neutrino mass matrix 
in the basis $\{(\nu_{iL})^C,\nu_{\alpha R},(S_{\rho L})^C \}$: 
\begin{eqnarray}
{\cal M}_\nu\ =\ \left(\begin{array}{ccc}
{\bf 0} & M_D & {\bf 0}\\
M_D^{\sf T} & \mu_R &  M_N^{\sf T}\\
{\bf 0} & M_N & \mu_S 
\end{array}\right)\; .
\label{eq:inverse1}
\end{eqnarray}
Note  that  we have  not  included  in  ${\cal L}_Y$  the  dimension-4
lepton-number  breaking term $\bar  L \Phi  S_L^C$ which  appears, for
instance, in  linear seesaw models~\cite{linear},  since the resulting
neutrino mass matrix in presence of this term can always be rotated to
the form given  in~(\ref{eq:inverse1})~\cite{marot}.  Observe that the
standard inverse  seesaw model discussed  originally in~\cite{inverse}
is recovered, once we set the  RH neutrino Majorana mass $\mu_R = {\bf
0}$  in  ~(\ref{eq:inverse1}).   The  smallness  of  the  $3\times  3$
symmetric  matrices $\mu_{R,S}$  are `technically  natural' in  the 't
Hooft sense. In  other words, in the limit  of $\mu_{R,S}\to {\bf 0}$,
lepton   number  symmetry   is  restored   and  the   light  neutrinos
$\nu_{1,2,3L}$  are massless  to all  orders in  perturbation, whereas
$\{\left(\nu_{1,2,3}\right)_R,\left(S_{1,2,3}\right)_L\}$  form  three
heavy     singlet    Dirac     neutrinos    of     approximate    mass
$\left(M_{N}\right)_{1,2,3}$.

In  this  paper,  we  analyze another  interesting  realization of 
inverse seesaw models,  where $\mu_R\neq {\bf 0}$, but  $\mu_S={\bf 0}$.  
In this case, the  light neutrinos are
massless at the  tree level, but acquire a small  mass at the one-loop
level. We  show that this one-loop  induced light neutrino
mass  is  {\em directly}  proportional  to  the  Majorana mass  matrix
$\mu_R$  and results  from  well-known SM  radiative corrections  that
involve the $Z$- and Higgs bosons~\cite{ap92}.  We call this scenario, the Minimal
Radiative Inverse Seesaw  Model. This is a rather economical scenario 
as it does not  require the existence
of  other  non-standard scalar  or  gauge  fields or other fermionic matter 
beyond the singlet neutrinos $\{\nu_{\alpha R},S_{\rho L}\}$.   
Finally, we  also
analyze  the  general  case,   where  both  $\mu_R$  and  $\mu_S$  are
non-vanishing  and present numerical  estimates regarding  the typical
relative size of  the two Majorana mass matrices  $\mu_R$ and $\mu_S$,
while maintaining agreement with neutrino oscillation data.

This paper is organized as follows: In Section II, we discuss the tree-level 
neutrino mass matrix in inverse seesaw models. In Section III, we present the one-loop neutrino masses and mixings in the Minimal Radiative Inverse Seesaw 
Model. In Section IV, we analyze the general inverse seesaw model at one-loop 
level. Finally, our conclusions are given in Section V. 

\section{Tree-Level Neutrino Masses} 

To obtain  the physical  neutrino masses, we  cast the mass  matrix in
(\ref{eq:inverse1})  into  a  form  similar to  the  type-I  seesaw case given 
in (\ref{eq:Mnutype1}) by defining a  $3\times 6$ mass matrix ${\cal
  M}_D=(M_D,{\bf 0})$  and a $6\times  6$ mass matrix ${\cal  M}_S$ in
the weak basis $\{(\nu_{1,2,3})_R,(S_{1,2,3})_L^C\}$:
\begin{equation}
	{\cal M}_S = \left(\begin{array}{cc}
\mu_R & M_N^{\sf T}\\
M_N & \mu_S
\end{array}\right)\; .
\label{eq:MS}
\end{equation}
We may now perform a block diagonalization of the $9\times 9$ neutrino
mass matrix~(\ref{eq:inverse1}) by a unitary transformation ${\cal V}$
to cast it into the form:
\begin{eqnarray}
{\cal V}^{\sf T} {\cal M}_\nu {\cal V} = \left(\begin{array}{cc}
M_{\nu_L} & {\bf 0}_{3\times 6} \\
{\bf 0}_{6\times 3} & {\cal M}_{\nu_R} 
\end{array}\right),
\label{eq:block}
\end{eqnarray}
where the  unitary matrix  ${\cal V}$ has  an exact  representation in
terms of a $3\times 6$ arbitrary matrix $\zeta$~\cite{ap93}:
\begin{eqnarray}
{\cal V} = \left(\begin{array}{ccc}
({\bf 1}_3+\zeta^*\zeta^{\sf T})^{-1/2} & \zeta^*({\bf 1_6}+\zeta^{\sf T}\zeta^*)^{-1/2}\\
-\zeta^{\sf T}({\bf 1}_3+\zeta^*\zeta^{\sf T})^{-1/2} & ({\bf 1}_6+\zeta^{\sf T}\zeta^*)^{-1/2}
\end{array}\right)\;,
\end{eqnarray} 
giving  rise to  the exact  analytic  expressions for  the block  mass
eigenmatrices~\cite{ap08}\footnote{Some approximate seesaw expressions are also studied in \cite{grim}.}:
\begin{eqnarray}
  \label{eq:mnulr1}
{\cal M}_{\nu_R} &=& ({\bf 1}_6+\zeta^\dag\zeta)^{-1/2}({\cal
  M}_S+{\cal M}_D^{\sf T}\zeta^*+\zeta^\dag {\cal M}_D)
({\bf 1}_6+\zeta^{\sf T}\zeta^*)^{-1/2}\; ,\\
  \label{eq:mnulr}
M_{\nu_L} &=& -({\bf 1}_3+\zeta\zeta^\dag)^{-1/2}({\cal
  M}_D\zeta^{\sf T}+\zeta {\cal M}_D^{\sf T}-\zeta {\cal M}_S\zeta^{\sf T})
({\bf 1}_3+\zeta^*\zeta^{\sf T})^{-1/2} = -\zeta {\cal
  M}_{\nu_R}\zeta^{\sf T}\;.
\end{eqnarray}

The  block diagonalization  condition implies  that the  $(1,2)$ block
element  of  ${\cal   V}^{\sf T}{\cal  M}_\nu{\cal  V}$  in~(\ref{eq:block})
vanishes, i.e.,
\begin{eqnarray}
  \label{eq:zeta}
{\cal M}_D-\zeta {\cal M}_S-\zeta{\cal M}_D^{\sf T}\zeta^* = {\bf 0}_{3\times 6},
\end{eqnarray}
which can  be solved for $\zeta$  in terms of ${\cal  M}_D$ and ${\cal
  M}_S^{-1}$. Moreover, ${\cal M}_S^{-1}$ is obtained by inverting the
mass matrix given in (\ref{eq:MS}):
\begin{widetext}
  \begin{eqnarray}
	{\cal M}_S^{-1} = \left(\begin{array}{ccc} 
		\left(\mu_R-M_N^{\sf T}\mu_S^{-1}M_N\right)^{-1} & 
		-\left(\mu_R-M_N^{\sf T}\mu_S^{-1}M_N\right)^{-1}M_N^{\sf T}\mu_S^{-1}\\
		-\left(\mu_S-M_N\mu_R^{-1}M_N^{\sf T}\right)^{-1}M_N\mu_R^{-1} & 
		\left(\mu_S-M_N\mu_R^{-1}M_N^{\sf T}\right)^{-1}
	\end{array}\right).
\label{eq:MSinv}
\end{eqnarray} 
\end{widetext}  
Note that  ${\cal M}_S^{-1}$  given by (\ref{eq:MSinv})  is symmetric,
since it is easy to show that for symmetric invertible matrices $\mu_{R,S}$,
\begin{eqnarray}
\left[\left(\mu_S-M_N\mu_R^{-1}M_N^{\sf
    T}\right)^{-1}M_N\mu_R^{-1}\right]^{\sf T} =\  
\left(\mu_R-M_N^{\sf T}\mu_S^{-1}M_N\right)^{-1}M_N^{\sf T}\mu_S^{-1}~.
\end{eqnarray}

\begin{figure}[h!]
\centering
\includegraphics[width=5cm]{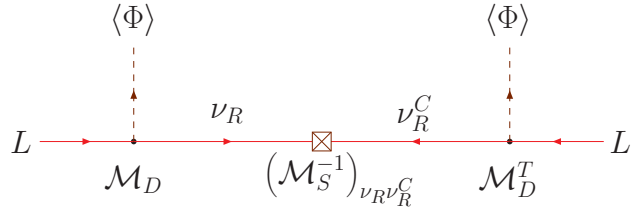}
\caption{The tree-level diagram for the light neutrino mass in the
  Inverse Seesaw Model.} 
\label{fig:1}
\end{figure}

The  physical   neutrino  mass  matrices   given  by~(\ref{eq:mnulr1})
and~(\ref{eq:mnulr}) can be expanded  in a converging Taylor series in
$\zeta$,  provided  its  norm   is  much  smaller  than  unity,  i.e.,
$||\zeta||\equiv  \sqrt{ {\rm Tr}(\zeta^\dag  \zeta)}\ll 1$,  which is
naturally satisfied  within the generic seesaw  framework.  To leading
order in  $||\zeta||$, (\ref{eq:zeta}) implies  $\zeta={\cal M}_D{\cal
  M}_S^{-1}$,  and   hence  the  light  neutrino   mass  matrix  given
by~(\ref{eq:mnulr}) simplifies to
\begin{eqnarray}
  \label{eq:mnulr2}
M_{\nu_L} &= & -{\cal M}_D{\cal M}_S^{-1}{\cal M}_D^{\sf T}+{\cal
  O}(||\zeta||^2)\nonumber\\ 
&=& -M_D\left({\cal M}_S^{-1}\right)_{\nu_R\nu_R^C}M_D^{\sf T}+{\cal
  O}(||\zeta||^2)\; .
\end{eqnarray}
This is analogous to the type-I seesaw formula~(\ref{eq:type1}), which
is  shown diagrammatically  in Figure~\ref{fig:1}.   Notice that  it is
only   the  $\nu_R\nu_R^C$   component  of   ${\cal   M}_S^{-1}$  that
contributes to the light neutrino mass matrix at the tree level.

In the limit $||\mu_{R,S}|| \ll  ||M_N||$, the $(1,1)$ block of ${\cal
M}_S^{-1}$  given by  (\ref{eq:MSinv}) can  be expanded  in  powers of
$\mu_S$.  We then obtain  from~(\ref{eq:mnulr2}) the  tree-level light
neutrino  mass matrix~\footnote{Notice  that $||\mu_{R,S}||$  could be
bigger than $||M_D||$ in this inverse seesaw approximation.}:
\begin{eqnarray}
M_{\nu_L}^{\rm tree} &=&  M_DM_N^{-1}\mu_S\left(M_N^{\sf T}\right)^{-1}M_D^{\sf T}
+ M_DM_N^{-1}\mu_S\left(M_N^{\sf T}\right)^{-1}\mu_RM_N^{-1}\mu_S
\left(M_N^{\sf T}\right)^{-1}M_D^{\sf T}
+ {\cal O}(\mu_S^3)\; , 
\label{eq:nul2}
\end{eqnarray}
which identically vanishes in the  limit $\mu_S\to {\bf 0}$.  This can
also be  deduced from the exact  expression in~(\ref{eq:mnulr}), since 
in this limit, $\zeta {\cal M}_D^{\sf T}={\bf 0}$ which implies 
$\zeta={\cal M}_D{\cal M}_S^{-1}$ from (\ref{eq:zeta}), 
and hence, $M_{\nu_L}={\bf 0}$ in 
(\ref{eq:mnulr}). 
Therefore, in this limit, the  full neutrino mass
matrix ${\cal M}_\nu$ in~(\ref{eq:inverse1})  has rank 6 and the light
neutrinos remain massless  at the tree level, even  if $\mu_R\neq {\bf
  0}$.

\section{Radiative Neutrino Masses and Mixings at the One-Loop Level}

\begin{figure}
\begin{center}
\includegraphics[width=5cm]{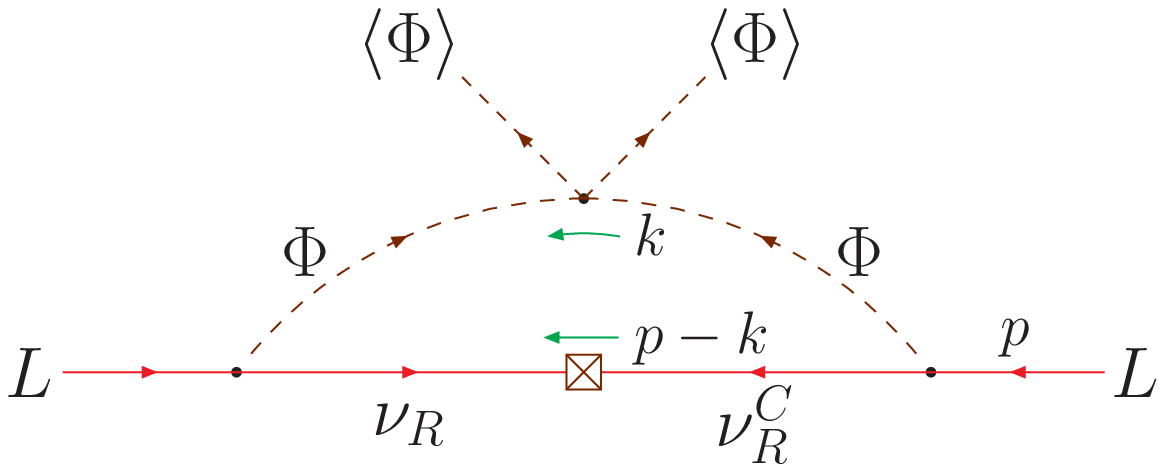}\\
\includegraphics[width=5cm]{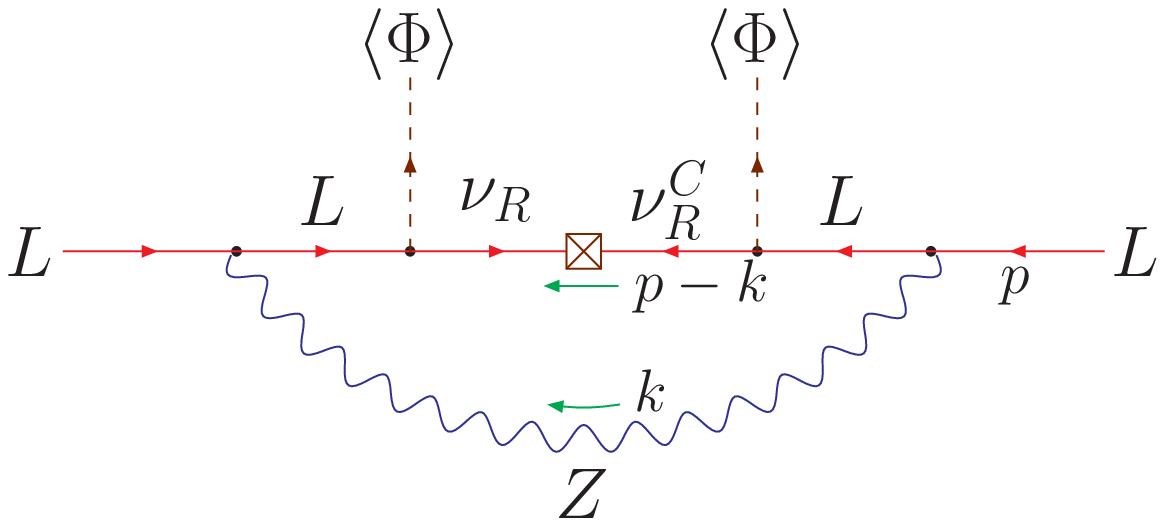}
\end{center}
\caption{One-loop  diagrams  pertinent  to  the  generation  of  light
  neutrino masses in the Minimal Radiative Inverse Seesaw Model.}
\label{fig:2}
\end{figure}

At  the one-loop  level, the  light neutrino  mass  matrix $M_{\nu_L}$
acquires a small radiative mass from the self-energy diagrams shown in
Figure~\ref{fig:2}.  We have drawn the  figures in the flavor basis to
explicitly show the lepton number violating mass insertions. Note here
that  the scalar  propagator $\Phi$  includes both  the  neutral Higgs
boson ($H$) and the neutral Goldstone boson ($G^0$) contributions.  In
the on-shell  renormalization scheme and  in the Feynman  gauge, these
are the only  one-loop diagrams contributing to the  neutrino mass, as
in  the type-I  seesaw case~\cite{ap92,Pilaftsis:2002nc,grim2}.   In a
general  $R_\xi$-gauge  parameterized  by the  gauge-fixing  parameter
$\xi_Z$, the  $\xi_Z$ dependence of  the self-energy for  the $Z$-loop
diagram is exactly  canceled by the $G^0$-loop diagram,  and the final
result is independent of $\xi_Z$~\cite{Pilaftsis:2002nc}.

The  one-loop induced  light  neutrino masses  are  obtained from  the
momentum-dependent self-energy  as follows:
\begin{eqnarray}
-M_{\nu_L}^{\rm 1-loop} P_R = \left.P_R\Sigma(\slashed p)P_R\right|_{\slashed p=0}~, 
\label{eq:self1}
\end{eqnarray}
where $P_R=(1+\gamma^5)/2$  is  the  right-chirality  projection
operator. Here $\Sigma(\slashed p)$  is the  sum  of the  self-energies of  the
diagrams  shown  in  Figure~\ref{fig:2},  and is given  by  the
following gauge-invariant Feynman amplitude:
\begin{eqnarray}
  \label{eq:self} 
P_R[i\Sigma(\slashed p)]P_R & =&  
\frac{g^2m_H^2}{4m_W^2}{\cal M}_D\int
\frac{d^dk}{(2\pi)^d}P_R\left[\frac{(\slashed p-\slashed k){\bf
      1}_6+{\cal M}_S}{(p-k)^2{\bf 1}_6-{\cal
      M}_S^2+i\epsilon}\right]\left[\frac{1}{(k^2-m_H^2+i\epsilon)^2}\right]
      P_R{\cal M}_D^{\sf T}\\
&&\hspace{-2.5cm} + 
\frac{g^2}{4\cos^2\theta_w}{\cal M}_D\int \frac{d^dk}{(2\pi)^d}P_R\gamma^\mu
\left[\frac{(\slashed p-\slashed
    k)}{(p-k)^2+i\epsilon}\right]\left[\frac{(\slashed p-\slashed
    k){\bf 1}_6+{\cal M}_S}{(p-k)^2{\bf 1}_6-{\cal
      M}_S^2+i\epsilon}\right]\left[\frac{(\slashed p-\slashed
    k)}{(p-k)^2+i\epsilon}\right]\gamma^\nu
P_R\left[\frac{g_{\mu\nu}}{k^2-m_Z^2+i\epsilon}\right]{\cal
  M}_D^{\sf T}\; ,\nonumber 
\end{eqnarray}
where $g$  is the weak gauge  coupling, $\theta_w$ is  the weak mixing
angle, and $d=4-\varepsilon$ is the dimensionality of space-time in the dimensional regularization scheme. Notice that the  integrands in~(\ref{eq:self})  vanish in
the infra-red (IR) limit $k^\mu\to 0$, because of the tree-level vanishing
condition on light neutrino masses.  However, the  loop  momentum $k$  is  integrated over  all
possible  values,  and  so  we  get a  non-zero  contribution  to  the
light-neutrino mass  matrix at the one-loop  level.  A straightforward
evaluation of  the ultra-violet (UV) finite Feynman  integrals appearing in
(\ref{eq:self}) leads to the following neutrino mass matrix in (\ref{eq:self1}):
\begin{eqnarray}
	M_{\nu_L}^{\rm 1-loop} &=&\frac{\alpha_W}{16\pi m_W^2}{\cal
          M}_D{\cal M}_S\left[\frac{m_H^2}{ {\cal M}_S^2-m_H^2{\bf
              1}_6}\ln\left(\frac{ {\cal M}_S^2}{m_H^2}\right) +  
	\frac{3m_Z^2}{ {\cal M}_S^2-m_Z^2{\bf 1}_6}\ln\left(\frac{
          {\cal M}_S^2}{m_Z^2}\right)\right]{\cal M}_D^{\sf T}\; , 
\label{eq:nuself}
\end{eqnarray} 
where   $\alpha_w\equiv   g^2/4\pi$  is   the   weak  fine   structure
constant.  In   the  limit   $||\mu_{R,S}||\ll  ||M_N||$  and   assuming  that
$M_N=m_N{\bf 1}_3$, for simplicity, we derive the simpler expression
\begin{eqnarray}
	M_{\nu_L}^{\rm 1-loop}\ \simeq\ \frac{f(x_N)}{m_W^2}\;
        M_D\mu_RM_D^{\sf T}\; ,
\label{eq:loop2}
\end{eqnarray}
where the one-loop function $f(x_N)$ is given by
\begin{eqnarray}
   \label{eq:loop}
f(x_N) &=& \frac{\alpha_W}{16\pi}
\left[\frac{x_H}{x_N-x_H}\ln\left(\frac{x_N}{x_H}\right)
	+\ \frac{3x_Z}{x_N-x_Z}\ln\left(\frac{x_N}{x_Z}\right)\right],
\end{eqnarray}
with  $x_N=  m_N^2/m_W^2$, $x_H=m_H^2/m_W^2$,  $x_Z=m_Z^2/m_W^2\equiv
1/\cos^2\theta_w$. From  (\ref{eq:loop2}), we clearly see  that at the
one-loop level, the light neutrino  mass depends {\em linearly} on the
lepton number breaking term $\mu_R$ (of dimension-3), and hence it does not 
vanish in the limit $\mu_S={\bf 0}$, unlike at the tree level.

\begin{figure}
\begin{center}
\includegraphics[width=8cm]{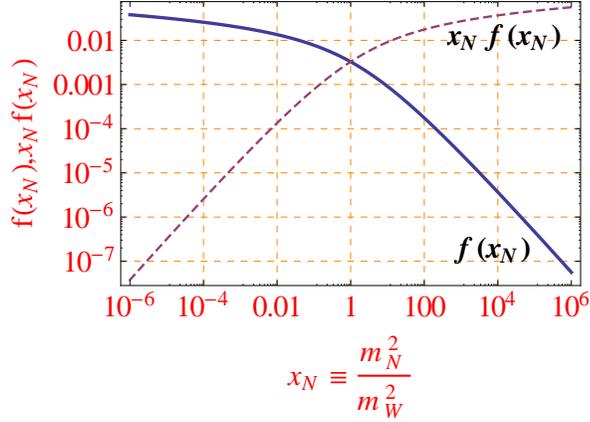}
\end{center}
\caption{The  analytic dependence  of the  one-loop  function $f(x_N)$
  [cf.~(\ref{eq:loop})] and  $x_N f(x_N)$  on the heavy  neutrino mass
  scale $x_N$.}
\label{fig:3}
\end{figure}

The  one-loop   function $f(x_N)$ given  by  (\ref{eq:loop})   is  plotted  in
Figure~\ref{fig:3},  for $x_H=1.88$  (corresponding  to $m_H=125$  GeV),
$x_Z=1.30$ and  $\alpha_W=0.033$.  Note that  in the limit  $x_N\gg 1$
(as  in the  canonical type-I  seesaw), the  one-loop  function stated
in~(\ref{eq:loop}) becomes
\begin{eqnarray}
f(x_N)\ \simeq\ \frac{\alpha_W}{16\pi}\;
\frac{x_H+3x_Z}{x_N}\ln(x_N)\; . 
\end{eqnarray}
Hence the light neutrino masses given by (\ref{eq:loop2}) are
suppressed by $1/m_N^2$: 
\begin{eqnarray}
M^{\rm 1-loop}_{\nu_L}\ \simeq\ \frac{\alpha_W}{16\pi}\;
\frac{M_D \mu_R M_D^{\sf T}}{m^2_N}\;
\frac{m_H^2+3m_Z^2}{m_W^2}\; \ln\left(\frac{m_N^2}{m_W^2}\right). 
\end{eqnarray}
On the other hand, it is
interesting to note that in  the limit $x_N\ll 1$, the one-loop function
takes on the form:
\begin{eqnarray}
f(x_N)\ \simeq\ \frac{\alpha_W}{16\pi}\;
\ln\left(\frac{x_Hx_Z^3}{x_N^4}\right)\; .
\end{eqnarray}
This  leads to  a light  neutrino mass  which is  only logarithmically
dependent on the heavy neutrino mass:
\begin{eqnarray}
M^{\rm 1-loop}_{\nu_L}\ \simeq\ \frac{\alpha_W}{16\pi}\; 
\frac{M_D\mu_R M_D^{\sf T}}{m^2_W}\;
\ln\left(\frac{m_Hm_Z^3}{m_N^4}\right)\;  . 
\end{eqnarray} 
It is  important to stress here that  this $\ln(m_N)$ dependence of the light 
neutrino mass is a unique  feature of the
Minimal Radiative Inverse Seesaw Model we have been studying here.

Finally,  we  wish to  offer  some comments  on  the  validity of  our
one-loop analysis presented here. In  a general study, one should also
consider the loop corrections to the other mass terms appearing in the
full  inverse seesaw mass  matrix in  (\ref{eq:inverse1}), as  well as
wave-function                        and                        mixing
renormalizations~\cite{ap96,Pilaftsis:2002nc}.    However,    in   the
$\overline{\rm MS}$ renormalization scheme,  the UV divergences of the
Dirac  mass in  the $\nu_L$--$\nu_R$  sector  can be  absorbed in  the
renormalized $M_D$.  On the other hand, all singlet Majorana and Dirac
mass  matrices   in  the  $\nu_{R}$--$S_L$  sector  do   not  get  any
UV-divergent contributions, at least at the one-loop level.

\section{The General Inverse Seesaw Model}

Let us  now consider the  general inverse seesaw scenario,  where both
the  Majorana  mass  matrices  $\mu_S$  and  $\mu_R$  do  not  vanish,
i.e.~$\mu_S\neq {\bf 0},\mu_R\neq {\bf 0}$.  To  leading order in $\mu_{R,S}$, the
light neutrino mass matrix is given by
\begin{eqnarray}
   \label{eq:gen}	
M_{\nu_L} &=&
        \frac{1}{m_N^2}M_D\Big(\mu_S+x_Nf(x_N)\mu_R\Big)M_D^{\sf T}
+{\cal O}(\mu_S^2,\mu_R^2)\; , 
\end{eqnarray}
where the one-loop function $f(x_N)$ is defined in~(\ref{eq:loop}). 

In order to get an order of magnitude estimate of the relative size of
$\mu_R$  with  respect  to   $\mu_S$  required  to  fit  the  neutrino
oscillation data,  let us choose a  basis in which  the charged lepton
mass matrix  is diagonal.  Then we may  write the light  neutrino mass
matrix as
\begin{eqnarray}
M_{\nu_L}\ =\ U^{\sf T}\widehat{M}_{\nu_L}^{\rm diag}U\; ,
\label{eq:mnulnum}
\end{eqnarray}
where $U$ is the PMNS mixing matrix\footnote{For simplicity, 
we have not considered here the leptonic non-unitarity effects which are 
generally of order $||\zeta||$ in inverse seesaw models~\cite{nuty}.} 
given in terms of the three
mixing angles, one Dirac and two Majorana $CP$ phases~\cite{vureview}: 
\begin{widetext}
\begin{eqnarray}  
U = \left(\begin{array}{ccc}
c_{12}c_{13} & s_{12}c_{13} & s_{13}e^{-i\delta}\\
-s_{12}c_{23}-c_{12}s_{23}s_{13}e^{i\delta} &
c_{12}c_{23}-s_{12}s_{23}s_{13}e^{i\delta} & s_{23}c_{13}\\ 
s_{12}s_{23}-c_{12}c_{23}s_{13}e^{i\delta} &
-c_{12}s_{23}-s_{12}c_{23}s_{13}e^{i\delta} & c_{23}c_{13} 
\end{array}\right)\times{\rm
  diag}(e^{i\alpha_1/2},e^{i\alpha_2/2},1)\; ,
\end{eqnarray}
\end{widetext}
with $c_{ij}\equiv \cos\theta_{ij},s_{ij}\equiv \sin\theta_{ij}$. 
For a normal hierarchy, one has
\begin{eqnarray}
\widehat{M}_{\nu_L}^{\rm diag} = {\rm diag}(m_1,\sqrt{\Delta m^2_{\rm
    sol}},\sqrt{\Delta m^2_{\rm atm}})\; . 
\end{eqnarray}
Using the central values of a recent global-fit analysis of the neutrino 
oscillation parameters~\cite{global}:
\begin{eqnarray}
\Delta m^2_{\rm sol}=  7.62\times 10^{-5}~{\rm eV}^2,~
\Delta m^2_{\rm atm} = 2.55\times 10^{-3}~{\rm eV}^2,~
\theta_{12}=34.4^\circ,~ \theta_{23}=40.8^\circ,\theta_{13}=9.0^\circ,
\delta=0.8\pi\; ,
\end{eqnarray}
and assuming $m_1=0,\alpha_1=\alpha_2=0$, we obtain from (\ref{eq:mnulnum}):
\begin{eqnarray}
	M_{\nu_L} &=& (10^{-2}~{\rm eV})\times \left(\begin{array}{ccc}
0.31-0.12i & -0.09+0.32i & -0.72+0.37i \\
-0.09+0.32i & 2.53+0.04i & 2.19+0.01i \\
-0.72+0.37i & 2.19+0.01i & 3.07-0.03i
\end{array}\right).
\end{eqnarray}

For  illustration, let  us absorb  the  flavor structure  of $M_D$  in
(\ref{eq:gen})  into  the   matrices  $\mu_{R,S}$,  by  rewriting  the
light-neutrino mass matrix as
\begin{eqnarray}
M_{\nu_L}\ =\ \lambda_D\,\Big(\bar\mu_S+x_Nf(x_N)\bar\mu_R\,\Big)\ \equiv\
\lambda_D\bar\mu_{\rm eff}\; , 
\label{eq:mnuplot}
\end{eqnarray}
where $\lambda_D= m_{D_{\rm max}}^2/m_N^2$ and $m_{D_{\rm max}}={\rm
max}\left(|(M_{D})_{ij}|\right)$ is the largest element of $M_D$ (in absolute-value
terms). In addition, we have re-defined $\mu_{S,R}$ as
\begin{eqnarray}
\bar{\mu}_S\ =\ K\mu_SK^{\sf T}\;, \qquad \bar{\mu}_R\ =\ K\mu_RK^{\sf T}\; ,
\end{eqnarray}
where $K=M_D/m_{D_{\rm max}}$ is in general a dimensionless $3\times
3$  complex  matrix.   With  these  definitions, we  can  now  make  a
one-to-one   mapping   between  the   elements   of  $M_{\nu_L}$   and
$\bar{\mu}_{R,S}$.
\begin{figure}[h!]
\begin{center}
\includegraphics[width=7cm]{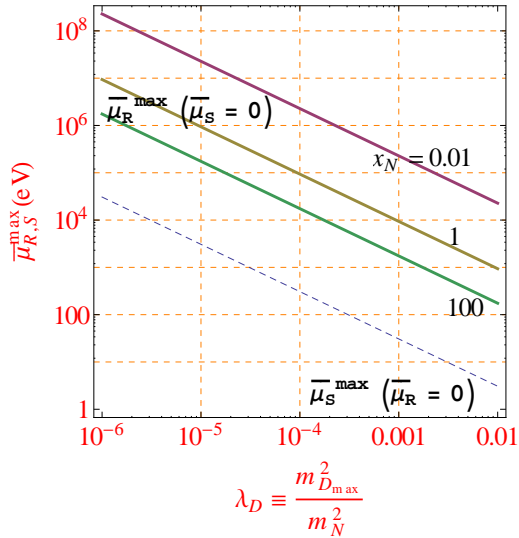}
\end{center}
\caption{The magnitudes of the largest element of $\bar\mu_S$ (dashed
  line) and $\bar\mu_R$ (solid lines for various values of $x_N$) as a
  function of the ratio $\lambda_D$.} 
\label{fig:mu}
\end{figure} 
This is shown  in Figure~\ref{fig:mu} where we have  plotted against the
ratio  $\lambda_D$:  (i)  the   magnitude  of  the  largest  entry  in
$\bar{\mu}_S$ (denoted as $\bar\mu_S^{\rm max}$) for $\bar\mu_R={\bf 0}$ (the  dashed line); (ii) the magnitude
of  the  largest entry  in  $\bar{\mu}_R$  (denoted as $\bar\mu_S^{\rm max}$) for  $\bar\mu_S={\bf 0}$ and  for
$x_N=0.01,1,100$ (the  solid lines).  We  note that the mass  scale of
$\mu_R^{\rm max}$ is  roughly $2$--$4$ orders  of magnitude larger than  that of
$\mu_S^{\rm max}$.   Thus, if  $\mu_R\neq {\bf 0}$,  a  much milder  hierarchy can  be
realized  between the  lepton-number  breaking scale  $\mu_R^{\rm max}$ and  
the electroweak scale $m_W$ in the Minimal Radiative Inverse Seesaw Model.

\begin{figure}[h!]
\begin{center}
\includegraphics[width=7cm]{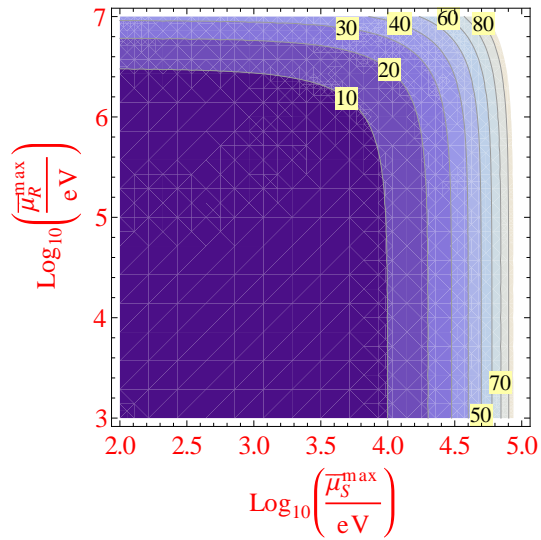}
\end{center}
\caption{Contours of constant $\bar\mu_{\rm eff}^{\rm max}$ (in keV)
  for $x_N=1$ and for $\bar\mu_{R,S}^{\rm max}>0$.} 
\label{fig:cont1}
\end{figure}

\begin{figure}[h!]
\begin{center}
\includegraphics[width=7cm]{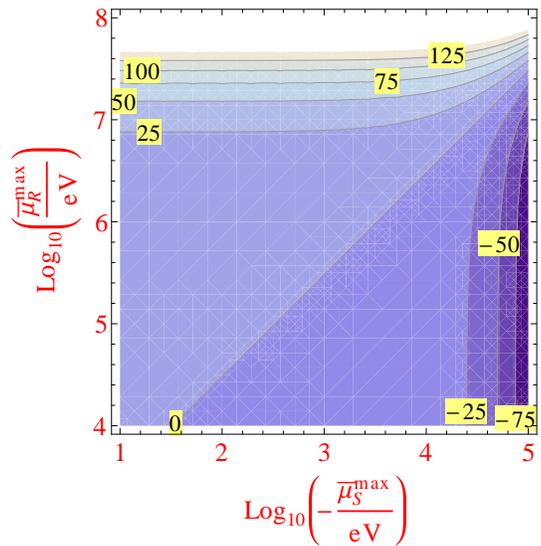}
\end{center}
\caption{Contours of constant $\bar\mu_{\rm eff}^{\rm max}$ (in keV)
  for $x_N=1$ and for $\bar\mu_{S}^{\rm max}<0,~\bar\mu_{R}^{\rm
    max}>0$.} 
\label{fig:cont2}
\end{figure}

In the  general inverse seesaw  scenario where $\mu_{R,S}\neq  {\bf 0}$, the
mass matrix $\bar\mu_{\rm eff}$ defined by (\ref{eq:mnuplot}):
\begin{eqnarray}
\bar\mu_{\rm eff}=\bar\mu_S+x_Nf(x_N)\bar\mu_R,
\end{eqnarray}
is fixed  by the  neutrino oscillation data  for a  given value of $\lambda_D$.
The dependence  of $\bar\mu_{\rm eff}^{\rm max}$  on $\lambda_D$ is  exactly 
the same as that of $\bar\mu_S^{\rm max}$ (with $\bar\mu_R={\bf 0}$), as shown by the dashed line
in Figure~\ref{fig:mu}.  However, the  relative size between $\mu_R$ and
$\mu_S$  depends on  their  relative  sign, for  a  fixed given  ratio
$x_N$. For instance, when both $\bar\mu_{S}^{\rm max},\bar\mu_{R}^{\rm
  max}>0$, the constant contours  of $\bar\mu_{\rm eff}^{\rm max}$ (in
keV)   are   shown   in   Figure~\ref{fig:cont1}   for   $x_N=1$.    For
$\bar\mu_{S}^{\rm   max}<0$   and   $\bar\mu_{R}^{\rm   max}>0$,   the
corresponding contours are shown in Figure~\ref{fig:cont2}, where we have 
the cancellation regions ($\bar\mu_{\rm eff}^{\rm max}\to 0$) for certain 
combinations of $\left(\bar\mu_S^{\rm max},\bar\mu_R^{\rm max}\right)$. 

We conclude this section by commenting that our numerical results only
depend on the largest element  of the light neutrino mass matrix which
is   always   of  the   order   of   $\sqrt{\Delta  m^2_{\rm   atm}}$,
irrespective  of the neutrino mass  hierarchy. Hence  our order  of magnitude
estimates will  be valid  for an inverted  hierarchical light neutrino
mass spectrum as well.

\section{Conclusions}

We have  presented a minimal radiative mechanism  for generating light
neutrino  masses  in inverse  seesaw  models.  The radiative  neutrino
masses arise at  the one-loop level from known  SM electroweak quantum
effects involving the $Z$ and  Higgs bosons. Unlike in other radiative
inverse seesaw mechanisms existing in the literature~\cite{radiative},
the  implementation of our  radiative mechanism  does not  require any
extra non-standard fields other than usual SM singlet neutrinos.

In the Minimal Radiative Inverse Seesaw Model, where the dimension-3 
lepton-number
breaking  mass  matrix  $\mu_R$   of  the  right-handed  neutrinos  is
non-zero, we  have found that  the light neutrino masses  generated at
one-loop level  are UV-finite and are 
{\em directly}  proportional to $\mu_R$.   We also
showed that  the 3-by-3  Majorana mass matrix  $\mu_R$ could  be $2-4$
orders of magnitude larger than  the other 3-by-3 Majorana mass matrix
$\mu_S$ present  in the standard  inverse seesaw models.   Hence, this
could alleviate the hierarchy between the lepton number breaking scale
and the electroweak scale in these models. In a supersymmetric version
of this mechanism (e.g.,~\cite{susy}), one might  expect to ameliorate this 
hierarchy even
further, in addition  to having a scalar dark  matter candidate in the
form of the lightest sneutrino~\cite{dm}. We hope to return to these issues in 
a future communication. 

\begin{acknowledgments}
This work is supported in part by the Lancaster-Manchester-Sheffield
Consortium for Fundamental Physics under STFC grant ST/J000418/1. In
addition, AP gratefully acknowledges partial support by a IPPP
associateship from Durham University.
\end{acknowledgments}

\end{document}